\documentclass[9pt,twocolumn,twoside]{IEEEtran}
\usepackage[utf8]{inputenc}
\usepackage[english]{babel}
\usepackage{authblk}
\usepackage{fancyhdr}
\usepackage{graphicx,xcolor}
\usepackage{amsmath,amsfonts,amssymb}
\usepackage[scaled]{helvet}
\usepackage[T1]{fontenc}
\usepackage{url}
\usepackage[colorlinks=true, allcolors=blue]{hyperref}
\usepackage{xifthen}
\usepackage{textcomp}
\usepackage{changepage}
\usepackage{cite}

\DeclareGraphicsExtensions{.pdf,.png,.jpg}

\title{Metacoatings for wavelength scale, high NA plano-concave focusing lenses}
\begin{document}
	
\author[1,2]{Mahin Naserpour}
\author[1,*]{Carlos J. Zapata-Rodr\'{i}guez}
\author[1]{Carlos D\'{\i}az-Avi\~n{\'o}}
\author[3]{Mahdieh Hashemi}
	
\affil[1]{Department of Optics and Optometry and Vision Science, University of Valencia, Dr. Moliner 50, Burjassot 46100, Spain}
\affil[2]{Department of Physics, College of Sciences, Shiraz University, Shiraz 71946-84795, Iran}
\affil[3]{Department of Physics, College of Science, Fasa University, Fasa 7461781189, Iran}
	
\affil[*]{Corresponding author: carlos.zapata@uv.es}
	
\date{Compiled \today}
	
	
	
\maketitle
\thispagestyle{fancy}
	
\begin{abstract}
	We design plano-concave silicon lenses with coupled gradient-index plasmonic metacoatings for ultrawide apertured focusing utilizing a reduced region of $\sim 20 \lambda^2$.
	The anomalous refraction induced in the planar input side of the lens and in the boundary of the wavelength-scale focal region boosts the curvature of the emerging wavefront, thus significantly enhancing the resolution of the tightly-focused optical wave. 
	The formation of a \textit{light tongue} with dimensions approaching those of the concave opening is here evidenced. 
	This scheme is expected to have potential applications in optical trapping and detection.
\end{abstract}
	

\section{Introduction}

Tight focusing of electromagnetic waves is essential to a myriad of applications such as photolithography and optical trapping, typically using dielectric thick lenses of high numerical aperture (NA).
However, the unceasing miniaturization and integrability of photonic devices imposes severe restrictions on the implementation of such sort of bulky pieces.
In this context, concave positive lenses composed of meta-atoms to change the sign of its macroscopic refractive index might be incorporated in high-NA focusing platforms within a reduced region \cite{Parazzoli04,Vodo05,Casse08,Xu15}. 
Epsilon-near-zero plano-concave lenses also prove a good performance prompted by the phenomenon of energy squeezing \cite{Beruete08,Navarro12}.
In particular, such architectures cannot operate at optical frequencies mainly due to material losses and demanding fabrication challenges. 

Advances in current techniques for nanofabrication, some of them based on plasmonics, have opened new prospects for producing compact, planar focusing systems. 
Plasmonic flat lenses enabling a plane-wave efficient shaping into an output converging wavefront within a subwavelength interaction space have been developed using arrays of nanoholes \cite{Gao10b,Ishii13}, nanoslits \cite{Ma10,Verslegers09b}, nanoparticles \cite{Ahmadi10,Huang14}, corrugations \cite{Wrobel09}, and nanoantennas \cite{Aieta12b,Yu13}.
In the previous examples, the mechanism set behind the control of the scattered light relies on the suitable spatial distribution of engineered, subwavelength phase shifters included on the active surface \cite{Epstein16}.
Therefore, a critical design aspect of the focusing device is the choice of the elementary scatterers for obtaining the required phase distribution. 

In the so-called metasurfaces, Mie resonances lead to controllable phase discontinuities which are applied to administrate the beam shaping \cite{Yu13}.
Alternatively, the phase of the electromagnetic field may change after passing through a subwavelength metal-dielectric structure on the basis of a modal coupling to the external radiation.
Such phenomenon is suitably explained in terms of an effective refractive index, and can be achieved with extremely-high performance in transmission \cite{Verslegers09}.
For instance, a set of metallic slits with controllable width can locally tune the effective index of refraction of the guided plasmonic modes \cite{Ishii11}.
This sort of ultrathin nanostructures are here coined as graded-index metacoatings.

Flat metalenses offer some advantageous accessibly to manufacture and integrate in complex devices, however suffer from severe geometrical restrictions to attain extremely-high numerical apertures \cite{Arbabi15,Byrnes16}.
An alternative design relies on sculpturing a concave surface.
In such manner an aplanatic metasurface patterned on a spherical substrate has been proposed to focus light without coma and spherical aberrations \cite{Aieta13}.
The concept of metamaterials with engineered dispersion in curvilinear coordinates, with coaxial and concentric geometries, has been extensively exploited for instance in hyperlensing \cite{Rho10,Barnakov11,Lu12}

Previously we presented on a design of non-planar metacoatings composed of subwavelength metal-dielectric arrays to accelerate optical waves in the near field \cite{Naserpour15,Naserpour15b,Naserpour15c}. 
These nanostructures allow to grade ultrahigh indices of refraction together with high transmissivity, demonstrating controllable phase and amplitude responses over subwavelength propagation ranges for TM-polarized waves. 
Here, we utilize such basic concept to engineer gradient-index ultrathin coatings, in a parallelizable assembly, as focusing elements with high efficiency.
In fact, advantages of cascading a set of metasurfaces have been previously reported for instance as a polarization rotator \cite{Huang15}, as an efficient cylindrical-vector-vortex beam generator \cite{Yi14}, and as a polarization-controlled lens \cite{Pfeiffer13}.
Conveniently implemented over the two sides of a plano-concave (divergent) dielectric microlens, we achieve miniaturized lenses with aberration-free high-numerical-aperture focalization.

\section{Lens design}

\begin{figure}[tb]
	\centering
	\fbox{\includegraphics[width=.95\linewidth]{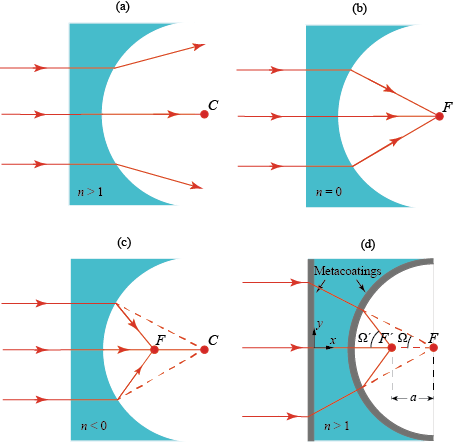}}
	\caption{
		Schematics based on optical rays of the focusing action of plano-concave dielectric lenses.
		(a) Transparent dielectrics with an index of refraction higher than unity lead to a diverging configuration.
		(b) An epsilon-near-zero metamaterial enables to focus light at the center of curvature of the concave surface.
		(c) An increased numerical aperture is attained by using negative-index metamaterials.  
		(d) Our proposal based on coupled metacoatings set at the entrance and exit surfaces of a transparent dielectric thick lens.
		A focused beam of semi-aperture angle $\Omega$ will be generated by passing through the gradient-index flat metasurface. 
		The converging wave field propagating inside the lens will be refocused at $F'$ by means of the active curved metacoating, having an increased semi-aperture angle $\Omega'$.
	}
	\label{fig01}
\end{figure}

The medullary component of our focusing device is a bulky dielectric lens which, without loss of generality, we consider as immersed in air.
Light impinges over the flat surface of the lens, and the exit spherical surface has a short radius $R > 0$, whose center is set at a given point $C$ as depicted in Fig.~\ref{fig01}.
Here we consider that $R$ remains in the scale of the wavelength, although our approach is not strictly restricted to such regime.
Under certain circumstances, this arrangement enables a high-aperture focusing of light emerging from the concave surface of the lens.
From the geometrical point of view, the optical power is carried by the second surface of the thick lens characterized by a paraxial focal length
\begin{equation}
	f = \frac{R}{1-n} ,
	\label{eq01}
\end{equation}
where $n$ denotes the refractive index of the lens.
If $n > 1$ as occurs in natural dielectric materials, the lens is divergent with negative focal length, $f < 0$, disabling to focus light behind the concave surface, as illustrated in Fig.~\ref{fig01}(a).
Therefore, focusing is only available with a refractive index below the unity, which can be achieved by using metamaterials \cite{Navarro12}.
In the specific case of epsilon-near-zero metamaterials, setting $n = 0$, light is focused at the center of curvature of the concave exit surface, as shown in Fig.~\ref{fig01}(b) \cite{Beruete08}.
This arrangement has recently been demonstrated experimentally in microwave frequencies and presents fascinating characteristics such as reduced aberrations in configurations with numerical apertures close to unity \cite{Torres15}.
An increased numerical aperture might be achieved by means of negative-index plano-concave lenses, as shown in Fig.~\ref{fig01}(c).
However, thick negative-index metamaterials in the near-infrared and visible frequency range exhibit a deficient performance mainly due to material Ohmic losses and inhomogeneities scattering.

Our approach is based on a transparent dielectric thick lens coated by graded-index ultrathin metasurfaces, as illustrated in Fig.~\ref{fig01}(d).
Therefore, material losses have a reduced impact in the lens performance.
The coated metasurfaces, here coined metacoatings, are designed to compensate the negative optical power of the plano-concave dielectric lens.
We propose a first metacoating set on the entrance flat surface to transform the incident plane wave into a spherical wave of center at $F$, coinciding with the center of curvature $C$ of the concave surface in the lens.
Such appropriate configuration reduces losses derived by reflections in the curved interface, and establishes a first numerical aperture in terms of the beam angle $\Omega$.
In order to even further increase the numerical aperture of the lens, a second metacoating is placed in the concave surface.
A controlled focal shift $a < R$ leads to a magnified far-field angle of the resulting converging wave \cite{Naserpour15d}.
As a result, the transmitted wave field will focus at $F'$ by free-space propagation, characterized by a numerical aperture $\mathrm{NA}' = \sin (\Omega')$.
Assuming that the focal point $F'$ of the coupled metacoating thick lens is located closer to the concave surface than $F$, that is $a = \overline{F F'} > 0$, the numerical aperture 
\begin{equation}
	\mathrm{NA}' = \frac{\mathrm{NA}}{\sqrt{1 + (a/R)^2 - 2 (a/R) \cos \Omega }}
	\label{eq04}
\end{equation}
is also higher than the numerical aperture $\mathrm{NA} = \sin  \Omega $ of the single-coated lens \cite{Naserpour15d}.
Specifically $\mathrm{NA}'$ reaches the unity when $a = R \cos \Omega$.
Note that the transverse and on-axis resolutions are dependent on the overall numerical aperture of the focusing lens.
In particular, this procedure leads to a critically improved axial resolution, which is determined by the inverse squared of the value of $\mathrm{NA}'$.

Next we design the gradient index metasurfaces which are overlaid on the lens surfaces, taking over the wavefront management of the incident plane wave.
Without loss of generality, we will consider cylindrical lenses enabling a description of the problem in two-dimensions.
Note that the dielectric material of the thick lens having a refractive index $n$ has a minor significance in the functioning of the focusing device.
Following the Fermat's principle by evaluating equal optical path lengths, a normally-incident plane wave passing through the first flat metacoating will be focused at the point $F$, found at a distance $f_1$ and transversally centered at $y = 0$, provided that the phase shift produced by the gradient-index metasurface yields:
\begin{equation}
	\phi_1 (y) = \phi_1 (0) + n k \left( f_1 - \sqrt{y^2 + f_1^2} \right),
	\label{eq02}
\end{equation}
where $k = 2 \pi / \lambda$ is the wavenumber in vacuum, $\phi_1 (0)$ is an arbitrary phase, and $y$ is the transverse coordinate indicating the distance from the optical axis.
Assuming an ultrathin metacoating of width $d$, the gradient index is then given by $n_1 (y) = \phi_1 (y) / k d$.
Of course, this approach disregards multiple reflections produced at the metacoating facets, and a more rigorous treatment will be implemented below.

As mentioned above, the cylindrical metacoating set at the back of the thick lens has a center of curvature $C$ that is located exactly at the focal point $F$.
Note that the constraint $R < f_1$ must be considered, where $f_1 - R$ stands for the vertex distance of the plane-concave lens.
Therefore the wavefront of the impinging field remains concentric to the concave metacoating arriving at normal incidence.
The induced phase shift on such metasurface will reshape the incident wavefront by increasing its curvature.
The field emerging from the nanolens then propagates in free space to focus at the point $F'$ that is shifted a length $a$ toward the rear metacoating.
The dephase induced by the cylindrical metasurface is \cite{Naserpour15d}
\begin{equation}
	\phi_2 (\theta) = \phi_2 (0) + k \left[ \left( R-a \right) - \sqrt{R^2 + a^2 -2 a R \cos \theta} \right] ,
	\label{eq03}
\end{equation}
where $\phi_2 (0)$ is again an arbitrary phase term, and $\theta$ is the azimuthal coordinate as measured from $C$ and determining the angle from the optical axis.
In practical terms, $\phi_2$ will be operative in the range $|\theta| \le \Omega$. 
If the width of the back metacoating is again $d$, the gradient index is now given by $n_2 (\theta) = \phi_2 (\theta) / k d$.

\begin{figure}[tb]
	\centering
	\fbox{\includegraphics[width=.95\linewidth]{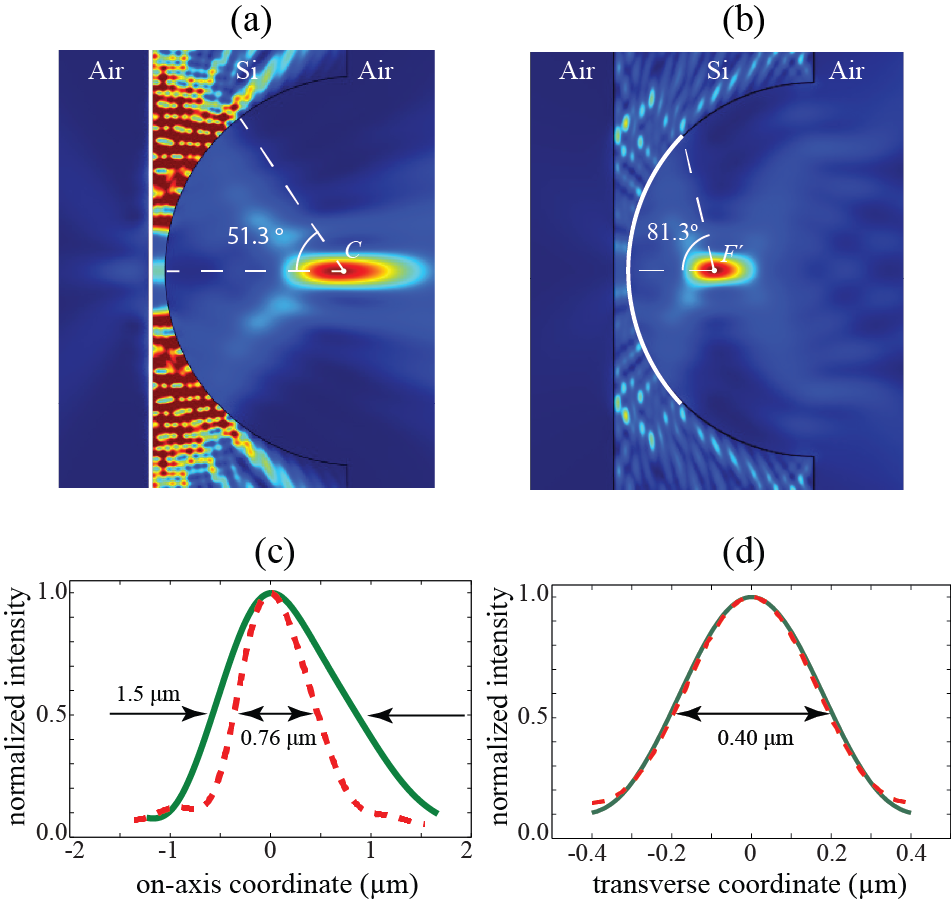}}
	\caption{
		(a) Intensity distribution $|H|^2$ generated by a nonuniform surface current with modulated phase distribution given by Eq.~(\ref{eq02}) and set at the front surface of a plano-concave Si lens (sketched in white solid line), mimicking the effect of the designer metacoating.
		In (b) we set the surface current with phase distribution given by Eq.~(\ref{eq03}) at the back surface of the Si lens.
		Normalized intensity of the magnetic field in the focal volume of the flat (cylindrical) surface current, represented in green solid lines (red dashed lines) as measured along (c) the $x$- and (d) the $y$- axis. 
		On axis resolution critically improves with an active cylindrical surface while transverse resolution does not change significantly.
	}
	\label{fig02}
\end{figure}

To illustrate the impact of ideal metacoatings, Fig.~\ref{fig02}(a) shows the intensity distribution of the magnetic field, $|H|^2$, produced by a nonuniform magnetic surface current set on the front lens facet and having a phase distribution governed by the expression given in Eq.~(\ref{eq02}).
The numerical simulations are calculated for a silicon plano-concave nanolens ($n = 3.69$) of radius $R = 3\ \mu$m at a wavelength $\lambda=800$~nm by using COMSOL Multiphysics, which is a finite element analysis software environment for modeling and simulation of electromagnetic systems.
In Fig.~\ref{fig02}(a), the length of the surface current is 8~$\mu$m, and it is designed to focus light at a distance $f_1 = 3.2\ \mu$m.
In this way we generate a focused field around the center of curvature $C$ with semi-aperture angle $\Omega = 51.3^{\circ}$. 
The effect of the back metacoating on the nanolens focal field is simulated in Fig.~\ref{fig02}(b), where we introduce a cylindrical surface current of curvature center set at $C$ and with semi-aperture angle $\Omega$.
Additionally, a phase distribution of the surface current following Eq.~(\ref{eq03}), for a focal shift $a = 1.5\ \mu$m, is implemented in the numerical simulation.
It is evident that the shaped focal field is axially shifted from $C$ to the focal point $F'$.
In addition, an increased semi-angular aperture $\Omega' = 81.3^{\circ}$, which is estimated with Eq.~(\ref{eq04}), will also make the numerical aperture NA' grow accordingly. 
Moreover, the enlarged numerical aperture leads to a resolution improvement in terms of the full width at half-maximum (FWHM) of the intensity, as measured along the $x$-axis, which is reduced from $1.5\ \mu$m to $0.76\ \mu$m as shown in Fig.~\ref{fig02}(c). 
Finally, the transverse resolution experiences a negligible improvement of 20~nm (the FWHM yields $0.40\ \mu$m with the back phased current) as evidenced in Fig.~\ref{fig02}(d), due to its proximity to the diffraction limit.

\section{Ultrathin graded-index metacoatings}

Next we proceed to design ultrathin metacoatings for the proper generation of the phase distribution $\phi_1(y)$ and $\phi_2(\theta)$, given in Eqs.~(\ref{eq02}) and (\ref{eq03}) respectively, all along the front surface and back surface of the plano-concave lens.
A great variety of strategies can be found in the literature for such a purpose, for instance gradient metasurfaces utilizing the Pancharatnam-Berry phase elements \cite{Lin14,Qin16} which in addition can provide a good efficiency.
However, plano-concave cylindrical lenses may benefit from simplicity in the fabrication of the active metacoating if for instance their elementary units should not be patterned along the cylinder axis.
A subwavelength metallic film of thickness $d \ll \lambda$ including nanoslits with controlled width proved to be good candidates for a tunable phase manipulation not only for spherical wavefronts \cite{Naserpour15d} but also to accelerate focal beams and to create light capsules \cite{Naserpour15,Naserpour15b,Naserpour15c}. 
The resulting metacoating can be considered as a graded-index uniaxial metamaterial by simply altering the metal filling fraction $f = w_m / (w_m + w_d)$ along the corresponding spatial coordinates, $y$ and $\theta$, where $w_d$ and $w_m$ stand for the width of the slit and the metallic segment, respectively.
Assuming that $w_m$ remains below the penetration depth of the wave field inside the metal, the periodic array of metallic slits with nanoscale size optically behaves like an effective uniaxial crystal \cite{Rytov56,Popov07}. 
When light passes through the metallic slits, coupled surface plasmons (SPs) are excited enabling wave propagation with characteristic propagation constant.
In these circumstances, the multilayered metal-dielectric material behaves like a semi-transparent effective medium exhibiting extreme anisotropy. 
Assuming that the incident plane wave is polarized perpendicularly to the nanoslits, the effective index of refraction of such metallodielectric metamaterial is approximately estimated as \cite{Yeh88}
\begin{equation}
	n_\mathrm{eff} = \sqrt{\frac{\varepsilon_d \varepsilon_m}{f \varepsilon_d + (1-f) \varepsilon_m}} ,
	\label{eq06}
\end{equation}
where $\varepsilon_m$ and $\varepsilon_d$ denote the relative permittivities of the metal and the medium in the slits, respectively.

\begin{figure}[tb]
	\centering
	\fbox{\includegraphics[width=.8\linewidth]{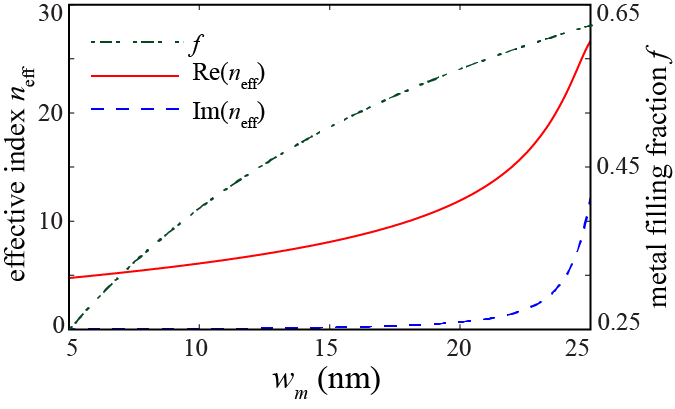}}
	\caption{
		Effective index of refraction evaluated with Eq.~(\ref{eq06}) for a gold-silicon periodic medium at a wavelength of $\lambda = 800$~nm.
		Silicon layers are set with a fixed width $w_d = 15$~nm.
		The metal filling fraction is governed by the Au films width $w_m$.
	}
	\label{fig03}
\end{figure}

As illustration, we show in Fig.~\ref{fig03} the effective index of refraction evaluated for a gold metamaterial patterned by a uniform distribution of nanoslits of constant thickness, $w_d = 15$~nm.
For a convenient design of the metamaterial, we consider that the nanoslits are filled with silicon, a material that is also employed for the bulky plano-concave lens.
Alternatively, one might propose a holey metacoating where the metallic slits are filled with air, but its lower index of refraction in comparison with Si critically limits the range of practical $n_\mathrm{eff}$.
At a wavelength  $\lambda = 800$~nm we set $\varepsilon_m = -23.36 + i 0.77$ for Au and $\varepsilon_d = 13.64$ ($ = n^2$) for Si.
An increasing metal filling fraction enables to establish an ultrahigh effective index $n_\mathrm{eff}$ that surpasses the refractive index of silicon.
However, this is attained at the cost of growing Ohmic losses.
For instance, a wave field propagating in an all-dielectric medium will gain a phase shift of $\pi/2$ radians if, instead, the material now includes metallic components of $w_m = 8.7$~nm ($\mathrm{Re} (n_\mathrm{eff}) = 5.69 = n + \lambda / 4 d$), assuming a propagation length as short as $d = 100$~nm.
Incremental phase shifts of $\pi/2$ radians can be obtained with Au elements of $w_m = 14.2$~nm and $w_m = 17.7$~nm, providing effective indices $\mathrm{Re} (n_\mathrm{eff}) = 7.69$ and $\mathrm{Re} (n_\mathrm{eff}) = 9.69$, respectively.

Note that the giant modulation of the effective index of refraction of our nanogratings cannot be experienced by plane waves with TE polarization.
In this case, the electric field is set along the nanoslits, and the real part of $n_\mathrm{eff}$ decreases when the metal filling factor grows \cite{Naserpour15b,Naserpour15d}.
For $w_m$ higher than 9~nm, the multilayered metamaterial behaves like a metal with $\mathrm{Re} (n_\mathrm{eff}) \approx 0$ and increasing $\mathrm{Im} (n_\mathrm{eff})$.
As a consequence, the designed metacoating is highly sensitive to the polarization of the incident light, and can only be used for TM-polarized incoming light.

\begin{figure}[tb]
	\centering
	\fbox{\includegraphics[width=.9\linewidth]{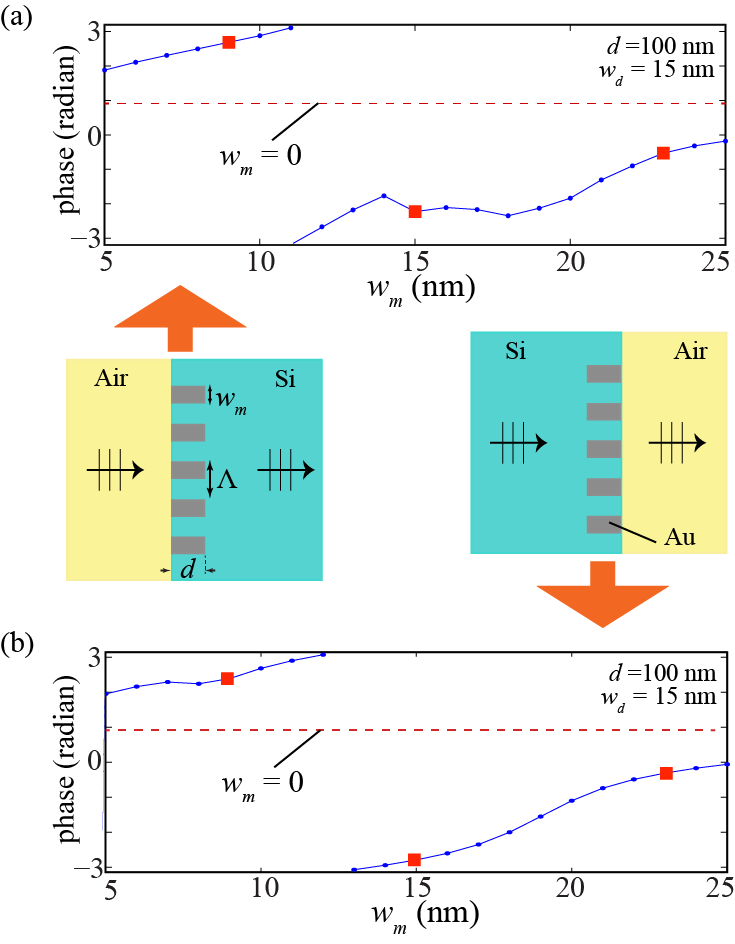}}
	\caption{
		Phase shift gained by a TM-polarized plane wave traversing through an Au-Si metacoating of thickness $d = 100$~nm, as set in an air/silicon plane interface.
		For simplicity, we represent the phase shift in an interval ranging from $-\pi$ to $\pi$.
		The slit width of the periodic nanostructure is fixed at $w_d = 15$~nm and we vary the metal width $w_{m}$.
		In (a) the beam impinges from air, and in (b) from silicon.
		The red dashed line establishes the phase shift measured for an all-dielectric coating ($w_m = 0$).
		Red squares illustrate that metamaterials with $w_m =$ 9, 15, and 23~nm producing incremental phase shifts of approximately $\pi /2$ radians with respect to a nonconducting film.
	}
	\label{fig04}
\end{figure}

The estimation of such rapidly-evolving phase shifts reached by wave fields in our Au-Si metamaterials is only approximate and more accurate evaluations are required to take into account for instance nonlocal effects \cite{Elser07}, setting some bounds in the homogenization of the metallodielectric medium, and also cavity resonances derived from multiple reflections in the metacoating interfaces \cite{Yeh88}, as we will detail below.
Fig.~\ref{fig04} shows the phase shift gained by a TM-polarized plane wave field of $\lambda = 800$~nm after passing through a gold film of width $d = 100$~nm including Si-filled nanoslits of thickness $w_d = 15$~nm.
The numerical simulations are again performed with COMSOL Multiphysics, which confirm (not shown in the figure) that the phase shift gained by the incident wave field is approximately proportional to the real part of $n_\mathrm{eff}$ as evaluated by Eq.~(\ref{eq06}), at least for metacoatings of moderate and low filling fraction where $w_m$ remains below 15~nm.
Fig.~\ref{fig04}(a) refers to the phase behavior of a nanopatterned film when beam impinges from air and propagates in Si after passing through such ultrathin device, as occurring in the front metacoating of a plano-concave lens.
The simulations reproducing the conditions of the metacoating at the back of the plano-concave lens are depicted in Fig.~\ref{fig04}(b).
Note that for a high metal filling fraction, we found that Eq.~(\ref{eq06}) overestimates the effective index of refraction of the metal-dielectric metamaterial. 
Nevertheless, we verify from our results (not entirely shown in the figure) that changing the parameter $w_m$, which determines the metal filling fraction of the nanostructure, enables a phase detuning of the transmitted optical signal within the required range of $2 \pi$ radians.

\begin{figure}[tb]
	\centering
	\fbox{\includegraphics[width=.8\linewidth]{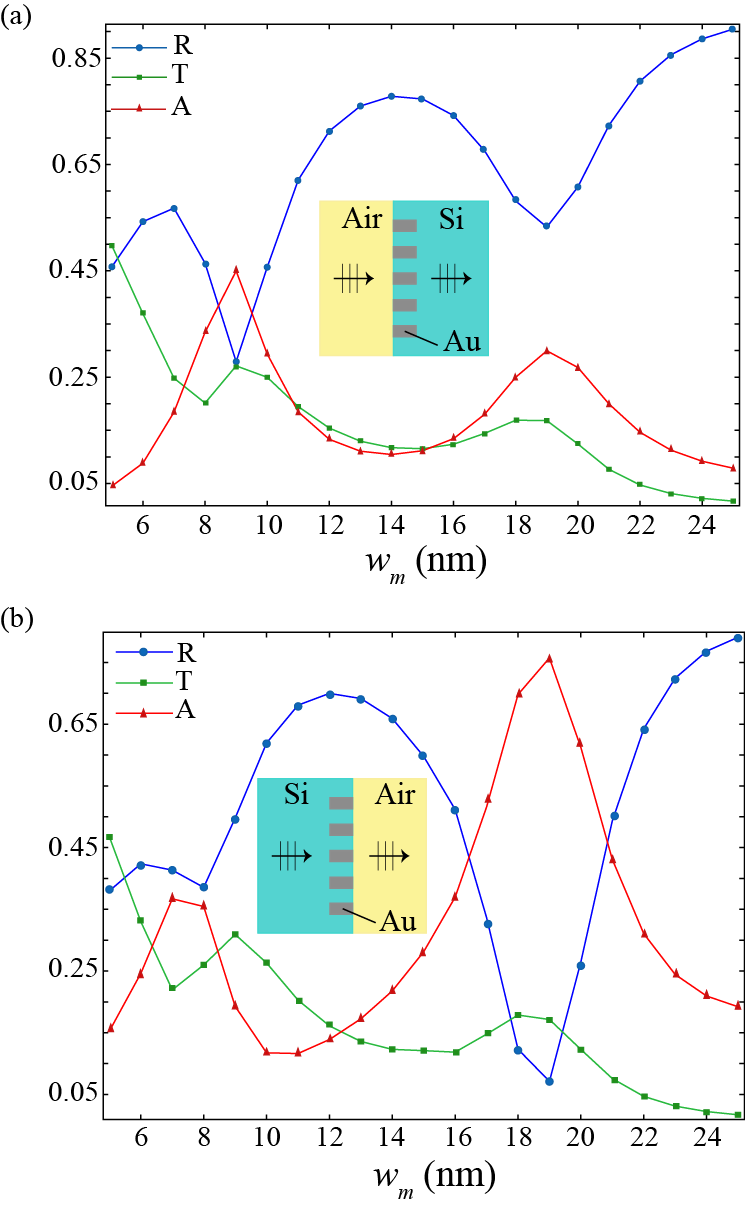}}
	\caption{
		Transmittance (T), reflectance (R) and absorptance (A) calculated for Au-Si metacoatings as described in Fig.~\ref{fig04}, varying the width $w_{m}$ of the metallic wires. 
		The beam impinges from (a) air and from (b) Si.
	}
	\label{fig05}
\end{figure}

Cavity resonances produced in the metacoating may lead to enhanced losses set in addition to Ohmic losses which are naturally present in the metal.
This effect is illustrated in Fig.~\ref{fig05}, where we evaluate the transmittance (also reflectance and absorptance) of our metacoatings of width $d = 100$~nm and nanoslit thickness $w_d = 15$~nm, employing our numerical solver based on the finite element method (FEM).
Fig.~\ref{fig05} evidences the relevance of absorption in the metamaterial for certain values of $w_m$ and also the loss of transmitted intensity driven by reflections.
Furthermore, nanostructures with values of $w_m$ higher than 25~nm are opaque in practical terms, a length which roughly represents the penetration depth in the metal.
We point out that surface plasmon resonances enabling extraordinary optical transmission, which is caused by the periodicity of the metallic nanostructure, are negligible in our case due to the deep subwavelength scale of $\Lambda = w_m + w_d$ \cite{Martin01}.
Now it is evident that a spatial modulation of the metal filling fraction in the metacoating leads not only to a graded-index $n_\mathrm{eff}$ which modifies the phase of the transmitted field but also contributes to the formation of gray zones where the intensity of the field can drop severely.

On the other hand, we may analytically estimate the amplitude and the phase transformation undergone by the scattered (magnetic) field after passing through the metallodielectric nanostructured film, by introducing the zero-order transmission amplitude:  \cite{Yeh88} 
\begin{equation}
	t = \frac{\tau_{12} \tau_{23} \exp(i k n_\mathrm{eff} d)}{1 - \rho_{21} \rho_{23} \exp(2 i k n_\mathrm{eff} d)} .
	\label{eq09}
\end{equation}
Here, $\tau_{12}$ and $\tau_{23}$ are transmission coefficients at the front and back surface of the metacoating, respectively, $\rho_{21}$ and $\rho_{23}$ are reflection coefficients at the two ends of the nanostructure, in all cases containing both magnitude and phase.
These coefficients are evaluated using $\tau_{\alpha \beta} = {n_\beta}/{(n_\alpha + n_\beta)}$
and $\rho_{\alpha \beta} = \tau_{\alpha \beta} - 1$, where $n_2 = n_\mathrm{eff}$.
Transmittance is simply computed as $T = |t|^2 (n_1 / n_3)$.
Then our metamaterial reaches its optimal transmittance when $\mathrm{arg} (\rho_{21} \rho_{23}) + 2 k d \mathrm{Re} (n_\mathrm{eff})$ satisfies the resonance condition reaching an entire multiple of $2 \pi$ radians.

From the analysis given above, it is apparent that an optimized response of the nanogratings in terms of the transmission efficiency might be achieved provided that the width of the Si layers in the multilayered nanostructure is not fixed but it can also vary appropriately. 
Nevertheless, further restrictions derived from fabrication tolerances in current nanotechnology support us to disregard narrower Si layers, enabling metamaterials with ultrahigh effective index of refraction, and also thinner metacoatings.

At this point we are in a position with the above-given analysis to design the elements of our graded-index metacoatings.
From a practical point of view, it is appropriate to perform a discrete phase shift rather than to apply a continuous modulation of the incident wavefront. 
In this case, a finite number of metal-dielectric nanostructures are necessary to obtain the outlined phase range. 
In particular, we provide four periodic metallodielectric nanostructures producing incremental phase shifts of $\pi / 2 $ radians, which has been proved to yield a satisfactory beam shaping \cite{Naserpour15b}.
We first consider an all-dielectric film ($w_m = 0$) which demonstrates an optimal transmittance.
In agreement with the results shown in Fig.~\ref{fig04}, the remaining three Au-Si slit composites with prescribed dephases of $\pi / 2 $ radians are then identified by $w_{m} = 9, 15, 23$~nm, respectively, when the silicon layer width is kept fixed as $w_d=15$~nm. 
These values are highlighted in red squares in Fig.~\ref{fig04}.
It is worth noting that current technology allows to assemble ultrathin, uniformly continuous metallic films of thicknesses of only $5\ \mathrm{nm}$ with reduced surface roughness of around $0.2\ \mathrm{nm}$ \cite{Chen10}.
As a consequence, the values of $w_m$ are here set as an integer in units of nm, and therefore the phase shift produced by these four elements of the graded-index metacoating are not exactly $\pi/2$~rads; nevertheless, the produced deviation is small enough to be neglected.

\begin{figure}[tb]
	\centering
	\fbox{\includegraphics[width=.95\linewidth]{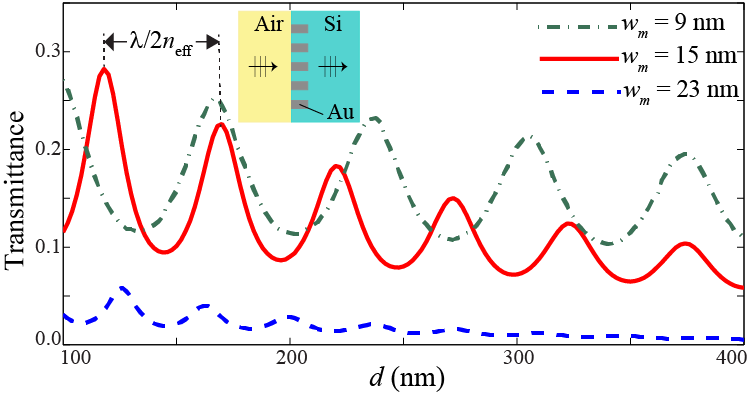}}
	\caption{Transmittance of metallic nanostructures with different Au wire width $w_m$ as a function of thickness $d$ of the metacoating, calculated at a wavelength $\lambda = 800$~nm.
		Here we set $w_d = 15$~nm.}
	\label{fig06}
\end{figure}

Finally we show the performance of our metallic gratings with different values of $d$.
In Fig.~\ref{fig06} we illustrate the high-to-moderate transparency of gratings with $w_{m} = 9,15,23$~nm and lengths $d$ ranging from $100$ to $400$~nm. 
As expected, Fabry-Pérot resonances with a periodic behavior governed by the effective index of refraction are responsible of the sinusoidal modulation of the transmitted field.
Specifically, here we verify that Eq.~(\ref{eq06}) clearly overestimates $n_\mathrm{eff}$ in the case of $w_m = 23$~nm.
Finally, we recognize that our approach does not permit an engineered phase modulation coinciding with peak transmittances through the elements of the metacoating, at least for short lengths $d$.

\section{Results and discussion}

Next we show some FEM-based numerical simulations to illustrate the validity of our approach, which relies on the use of coupled metacoatings set on the front and back surfaces of a wavelength scale plano-concave lens thus enabling a plane wave to be tightly focused. 
As detailed above, the first metacoating is designed to convert a plane wavefront into a cylindrical wavefront that is concentric to the back surface at the point $C$, and the second metasurface further increases the angular aperture of the miniaturized lens.
The metacoatings are composed of four types of zones, three of them constituted by metal-dielectric subwavelength gratings to induce controlled phase shifts in multiples of $\pi/2$ with respect to bulk Si.
We considered nanopatterned films of thickness $d = 100$~nm with slits of $w_d = 15$~nm, arranged in arrays of period $\Lambda_{1}=24$~nm, $\Lambda_{2}=30$~nm, and $\Lambda_{3}=38$~nm. 
The phase pattern of the transmitted fields, approaching the continuous distribution given in Eqs.~(\ref{eq02}) and (\ref{eq03}), are symmetric with respect to the center of each metacoating, and therefore the designed metacoatings exhibit a mirror symmetry with respect to $y = 0$.

\begin{figure}[tb]
	\centering
    \fbox{\includegraphics[width=0.9\columnwidth]{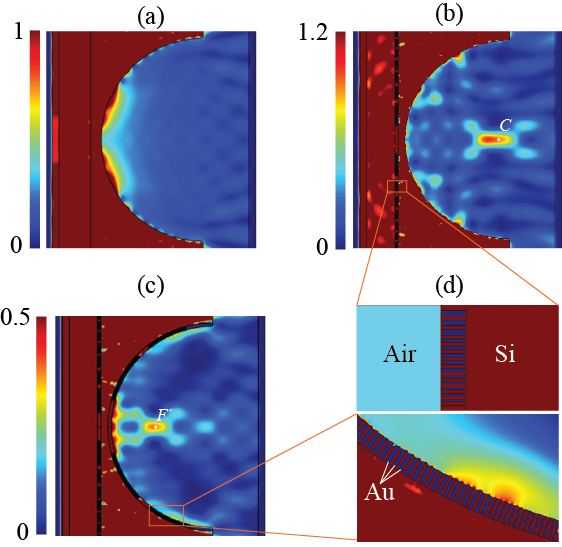}}
	\caption{
		FEM-based numerical simulations showing the intensity of the magnetic field when a monochromatic TM-polarized plane wave passes through a Si plano-concave lens of radius $R = 3\ \mu$m and vertex distance of 200~nm:  
		(a) without metacoatings,  
		(b) including a single metacoating set on the flat front surface, and
		(c) with coupled metacoatings lying on the front and back surfaces of the lens.
		(d) Close up of patterned Au nanoslit arrays in the flat (top) and concave (bottom) surfaces of the Si lens.
	} 
	\label{fig07} 
\end{figure}

Fig.~\ref{fig07}(a) shows the refractive behavior of a bare Si plano-concave lens of radius $R = 3\ \mu$m and vertex distance of 200~nm when a TM-polarized plane wave is used with $\lambda = 800$~nm.
Note that the effective area of the thick lens including the concave focal region is roughly $A_\mathrm{eff} = 2 R f_1$, which represents an extent of only $30 \lambda^2$. 
In agreement with Fig.~\ref{fig01}(a), the beam spreads out by passing through the thick lens and it is unable to focus light.
Figure~\ref{fig07}(b) shows the performance of the same Si plano-concave lens but including a flat metacoating in its front surface to focus the beam at a distance of $3.2\ \mu$m from it, exactly at the center point $C$. 
In particular, the central element of the metacoating (Si) has a semi-length of 400~nm, followed by a nanograting of period $\Lambda_3$ and total length of 304~nm (8~periods), an elementary grating of period $\Lambda_2$ and 210~nm-length (7~periods), and a metal-dielectric periodic nanostructure of period $\Lambda_1$ and length of 192~nm (8~periods); this sequence of elementary gratings is repeated to gain the necessary phase shift $\phi_1 (y)$ given in Eq.~(\ref{eq02}).
Comparing with Fig.~\ref{fig02}(a), one can realize that the result of using elementary gratings to induce phase shifts of $\pi/2$ radians is in a good agreement with the optimal designer metacoating producing a continuous phase distribution. 
Finally, we combine the action of a flat metacoating and a cylindrical metacoating set simultaneously, as shown in Fig.~\ref{fig07}(c), thus obtaining a focal shift of $a = 1.5\ \mu$m and consequently a super-resolved focal spot. 
The arched metacoating has a central silicon piece of apical semi-angle $15^\circ$, assembled to a periodic nanostructure of Au filling factor $f = 0.61$ and aperture angle $\Delta \theta = 11.7^\circ$ (16 periods), a cylindrical nanograting of $f = 0.50$ and angle $9.12^\circ$ (15 periods), a curved metamaterial of $f = 0.38$ and angular range of $7.9^\circ$ (17 periods), and so on to achieve the designed phase shift $\phi_2 (\theta)$.
In particular, the on-axis FWHM of the field intensity is decreased from $1.12\ \mu$m for the case shown in Fig.~\ref{fig07}(b) to $0.84\ \mu$m measured from Fig.~\ref{fig07}(c).

Here we point out that in spite of the low transmittance experienced by the elementary grating with $w_m = 23$~nm in comparison with other metal-dielectric nanostructures, such phase shifter is still applicable.
The modulation undergone by the real amplitude of the scattered field has minimal effect regarding the construction of the focused wave, which might remind a Fresnel zone plate. 
It affects the energy efficiency of the lens and perhaps in the emergence of sidelobes and secondary peaks. 
It can be seen by comparing the resulting focal field in Fig.~\ref{fig02}(b), where no amplitude modulation is applied, and Fig.~\ref{fig07}(c). 

\begin{figure}[tb]
	\centering
    \fbox{\includegraphics[width=.8\columnwidth]{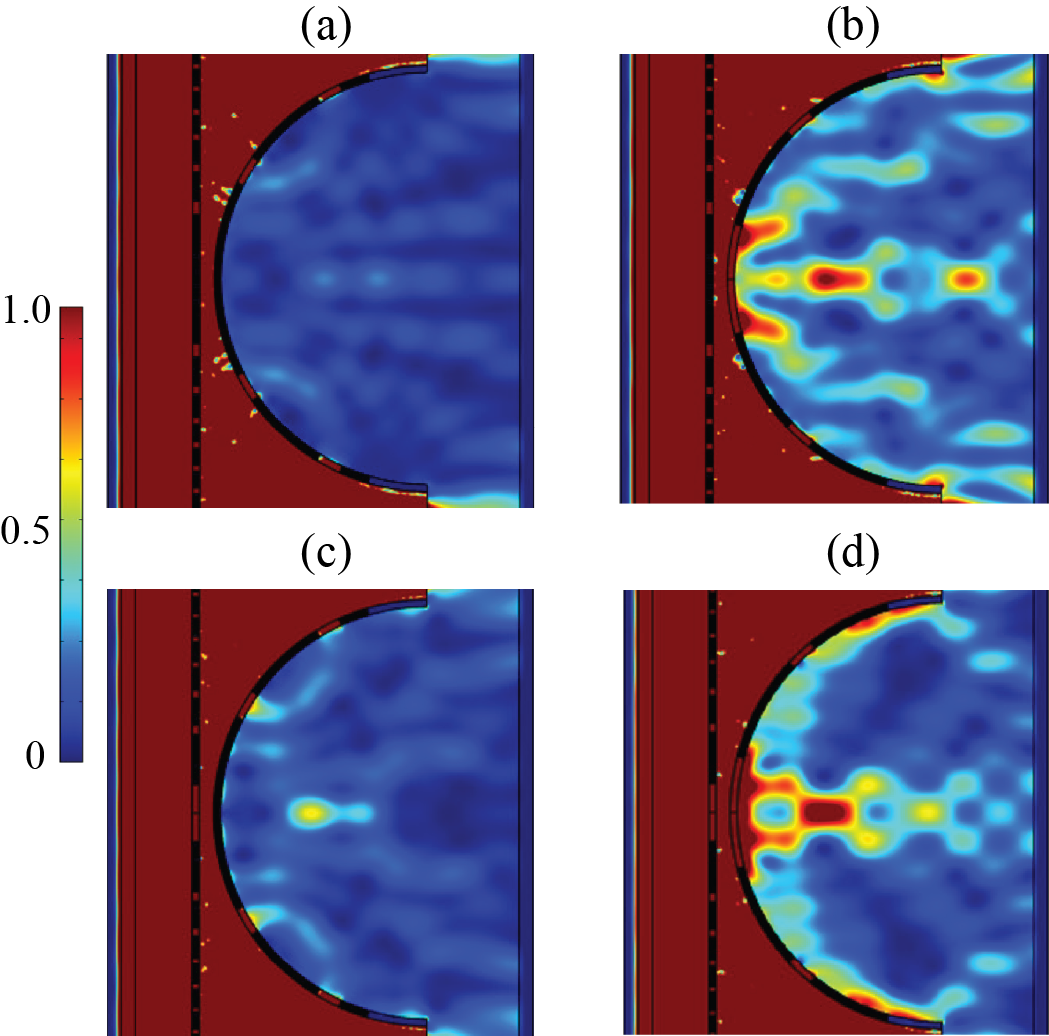}}
	\caption{
		Intensity distribution produced in the focal region of a Si plano-concave lens including metacoatings with different arrangements of elementary metal-dielectric gratings. 
		(a) A metallic grating of period $\Lambda_3 = 38$~nm is used at the central zone of both metacoatings. 
		Alternatively, we use an all-dielectric central zone for one metacoating and a metallic grating of period $\Lambda_3$ in the center of: (b) the cylindrical metacoating, and (c) the front flat metacoating.
		In (d) we reproduce Fig.~\ref{fig07}(c), where the central zone of both metacoatings has no metallic components, but here using the same colormap of previous sub-figures.
	} 
	\label{fig08} 
\end{figure}

Note that the metacoatings used in Fig.~\ref{fig07} are engineered to provide the maximum energy density at focus $F'$, a condition that is satisfied when the metal-dielectric gratings are symmetrically set off axis.
In this case, the central zone around $y = 0$ shaping a major energy flux of the resultant focused field presents reduced losses derived by reflection and absorption, as shown in Fig.~\ref{fig05}.
To illustrate such effect, in Fig.~\ref{fig08} we consider metacoatings whose central zone is formed by a metal-dielectric grating of period $\Lambda=38$~nm, which presents dramatic losses due to its high metal filling factor.
In order to obtain the required phase shift described in Eqs.~(\ref{eq02}) and (\ref{eq03}), the distribution of elementary metallodielectric gratings should be permuted accordingly.
When such lossy nanostructure is set in the central zone of the front metacoating and the back metacoating simultaneously, the plano-concave focusing lens is essentially opaque, as shown in Fig.~\ref{fig08}(a).
The impact of setting such lossy grating in the central zone is much notable on the cylindrical back metacoating rather than in the flat front metacoating, as depicted in Figs.~\ref{fig08}(b) and (c).
In the latter case, however, spurious spots appear in the focal volume, with special relevance of that located at the central point $C$.
Finally, the noisy focal waves are clearly corrected when semi-transparent phase shifters are set in the central zone of the coupled metacoatings, as shown in Figs.~\ref{fig08}(d).

\begin{figure}[tb]
	\centering
    \fbox{\includegraphics[width=.95\columnwidth]{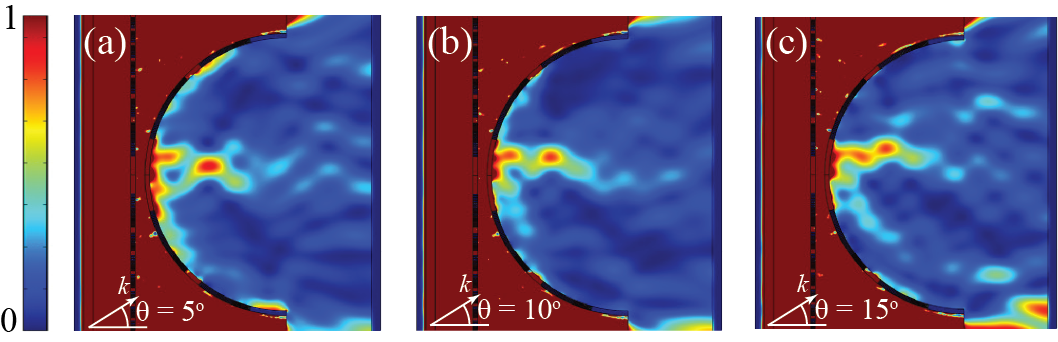}}
	\caption{Intensity distribution of focal waves produced by tilted TM-polarized plane waves with angles (a) $\theta=5^{\circ}$, (b) $10^{\circ}$, and (c) $15^{\circ}$, all measured with respect to the optical axis $y = 0$.} 
	\label{fig09} 
\end{figure}

Our plano-concave lens with coupled metacoatings also presents a favorable performance under oblique incident illumination.
For a comprehensive analysis of the refraction (and reflection) phenomenon involved in the interaction of monochromatic beams entering in metal-dielectric multilayered metamaterials see Ref.~\cite{Diaz16}.
As shown in Fig.~\ref{fig09}, the beam is not focused at the focal point $F'$ but the spot is laterally displaced as long as the impinging plane wave is tilted for paraxial angles ranging from $\theta = 5^{\circ}$ to $\theta = 15^{\circ}$.
The resulting focal wave does not present a significant atrophy of the beam shape and, consequently, the lateral and on-axis resolution remain practically unaltered.

\begin{figure}[tb]
	\centering
    \fbox{\includegraphics[width=.7\columnwidth]{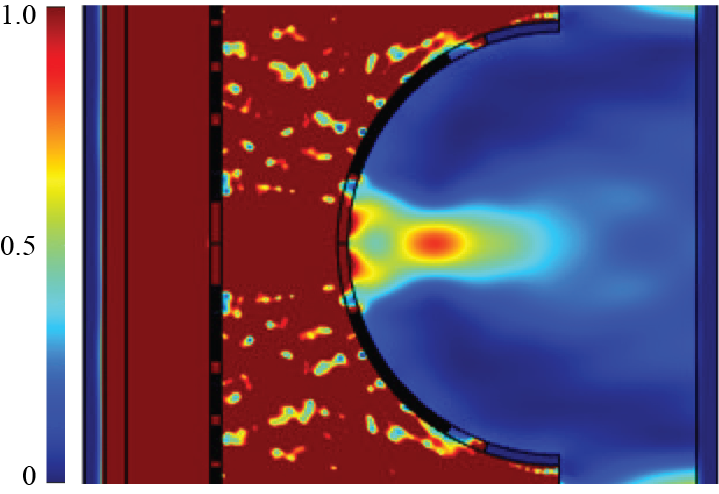}}
	\caption{Intensity of the magnetic field in the focal region of a metacoated Si plano-concave lens of radius $R = 2\ \mu$m, setting the focal shift parameter as $a = 1\ \mu$m.} 
	\label{fig10} 
\end{figure}

In specific applications such as particle trapping and sensing, it might be of interest to use high-numerical-aperture plano-concave lenses in the scale of the wavelength where $R \gtrsim \lambda$.
In this case, the concave opening operates like an \textit{optical mouth} within which it is possible to observe the formation of a focal wave in the form of a \textit{light tongue} with appropriate dimensions.
Here we explore the behavior of our proposal as pushed to such a limit. 
In Fig.~\ref{fig10} we show the intensity of a focused wave field performed by a metacoated plano-concave lens of inner radius $R = 2\ \mu$m where the focal point shifted from $C$ a distance $a = 1\ \mu$m, again using a working wavelength of $\lambda = 800$~nm.
Since $f_1$ is kept fixed for simplicity, the thickness of the Si lens increased from $0.2\ \mu$m to $1.1\ \mu$m, which carries no impact in the generation of the focal wave.
Here the effective area of the thick lens including the focal region decreases to $A_\mathrm{eff} = 20 \lambda^2$.
As suggested above, an all-dielectric central zone is used to maximize the energy efficiency of the focusing device.
The resultant focal spot again remains near the diffraction limit without a significant distortion in comparison with a virtually aberration-free focal wave.
We may conclude that the reduction of the lens radius can be proceeded in effect however limited by diffraction.

\section{Conclusions}

We designed cascaded metacoatings for the plano-concave surfaces of a wavelength-scale silicon lens to tightly focus light in the interior of the hollow opening in the form of a light tongue, which can shift laterally by controlling the tilt angle of the incident plane wave.
The gradient-index metacoatings are here composed of a patterned gold film with engineered nanoslits, thus offering reduced Ohmic losses and efficient coupling to the exterior radiation.
We point out that parallelization of metacoatings and alternatively metasurfaces might be carried out by means of auxiliary phase shifters like nanoholes and nanoantennas.
Furthermore, suitable designs for terahertz and lower frequencies can also benefit from engineered metamaterials with ultrahigh index of refraction \cite{Shin09,Choi11,Lee15}.
Mid-IR wavefront shaping can be carried out also using all-dielectric subwavelength high-index-contrast gratings offering a low-loss performance \cite{Huang16}.
Finally, compact multifocal nanolenses might be developed following for instance the procedure of Refs.~\cite{Mudry10,Hashemi16}.
Implemented in a lenslet array, multifunctional platforms might be conceived for potential applications in sensing and optical trapping, where the light tongue is optically activated.

\textbf{Funding.} Spanish Ministry of Economy and Competitiveness (MEC) under contract TEC2014-53727-C2-1-R.


\begin{thebibliography}{10}
	\newcommand{\enquote}[1]{``#1''}
	
	\bibitem{Parazzoli04}
	C.~G. Parazzoli, R.~B. Greegor, J.~A. Nielsen, M.~A. Thompson, K.~Li, A.~M.
	Vetter, M.~H. Tanielian, and D.~C. Vier, \enquote{Performance of a negative
		index of refraction lens,} Appl. Phys. Lett. \textbf{84}, 3232--3234 (2004).
	
	\bibitem{Vodo05}
	P.~Vodo, P.~V. Parimi, W.~T. Lu, and S.~Sridhar, \enquote{Focusing by
		planoconcave lens using negative refraction,} Appl. Phys. Lett. \textbf{86},
	201108 (2005).
	
	\bibitem{Casse08}
	B.~D.~F. Casse, W.~T. Lu, Y.~J. Huang, and S.~Sridhar, \enquote{Nano-optical
		microlens with ultrashort focal length using negative refraction,} Appl.
	Phys. Lett. \textbf{93}, 053111 (2008).
	
	\bibitem{Xu15}
	J.~Xu, Y.~Zhong, S.~Wang, Y.~Lu, H.~Wan, J.~Jiang, and J.~Wang, \enquote{Focus
		modulation of cylindrical vector beams by using {1D} photonic crystal lens
		with negative refraction effect,} Opt. Express \textbf{23}, 26978--26985
	(2015).
	
	\bibitem{Beruete08}
	M.~Beruete, M.~Navarro-C\'{\i}a, M.~Sorolla, and I.~Campillo,
	\enquote{Planoconcave lens by negative refraction of stacked subwavelength
		hole arrays,} Opt. Express \textbf{16}, 9677--9683 (2008).
	
	\bibitem{Navarro12}
	M.~Navarro-C\'{\i}a, M.~Beruete, M.~Sorolla, and N.~Engheta, \enquote{Lensing
		system and {Fourier} transformation using epsilon-near-zero metamaterials,}
	Phys. Rev. B \textbf{86,}, 165130 (2012).
	
	\bibitem{Gao10b}
	H.~Gao, J.~K. Hyun, M.~H. Lee, J.-C. Yang, L.~J. Lauhon, and T.~W. Odom,
	\enquote{Broadband plasmonic microlenses based on patches of nanoholes,} Nano
	Lett. \textbf{10}, 4111--4116 (2010).
	
	\bibitem{Ishii13}
	S.~Ishii, V.~M. Shalaev, and A.~V. Kildishev, \enquote{Holey-metal lenses:
		Sieving single modes with proper phases,} Nano Lett. \textbf{13}, 159--163
	(2013).
	
	\bibitem{Ma10}
	C.~Ma and Z.~Liua, \enquote{A super resolution metalens with phase compensation
		mechanism,} Appl. Phys. Lett. \textbf{96}, 183103 (2010).
	
	\bibitem{Verslegers09b}
	L.~Verslegers, P.~B. Catrysse, Z.~Yu, J.~S. White, E.~S. Barnard, M.~L.
	Brongersma, and S.~Fan, \enquote{Planar lenses based on nanoscale slit arrays
		in a metallic film,} Nano Lett. \textbf{9}, 235--238 (2009).
	
	\bibitem{Ahmadi10}
	A.~Ahmadi, S.~Ghadarghadr, and H.~Mosallaei, \enquote{An optical reflectarray
		nanoantenna: The concept and design,} Opt. Express \textbf{18}, 123--133
	(2010).
	
	\bibitem{Huang14}
	Y.~Huang, Q.~Zhao, S.~K. Kalyoncu, R.~Torun, Y.~Lu, F.~Capolino, and O.~Boyraz,
	\enquote{Phase-gradient gap-plasmon metasurface based blazed grating for real
		time dispersive imaging,} Appl. Phys. Lett. \textbf{104}, 161106 (2014).
	
	\bibitem{Wrobel09}
	P.~Wr\'obel, J.~Pniewski, T.~J. Antosiewicz, and T.~Szoplik, \enquote{Focusing
		radially polarized light by a concentrically corrugated silver film without a
		hole,} Phys. Rev. Lett. \textbf{102}, 183902 (2009).
	
	\bibitem{Aieta12b}
	F.~Aieta, P.~Genevet, M.~A. Kats, N.~Yu, R.~Blanchard, Z.~Gaburro, and
	F.~Capasso, \enquote{Aberration-free ultrathin flat lenses and axicons at
		telecom wavelengths based on plasmonic metasurfaces,} Nano Lett. \textbf{12},
	4932--4936 (2012).
	
	\bibitem{Yu13}
	N.~Yu, P.~Genevet, F.~Aieta, M.~A. Kats, R.~Blanchard, G.~Aoust, J.-P.
	Tetienne, Z.~Gaburro, and F.~Capasso, \enquote{Flat optics: Controlling
		wavefronts with optical antenna metasurfaces,} IEEE J. Sel. Top. Quantum
	Electron. \textbf{19}, 4700423--4700423 (2013).
	
	\bibitem{Epstein16}
	A.~Epstein and G.~V. Eleftheriades, \enquote{Huygens’ metasurfaces via the
		equivalence principle: design and applications,} J. Opt. Soc. Am. B
	\textbf{33}, A31--A50 (2016).
	
	\bibitem{Verslegers09}
	L.~Verslegers, P.~B. Catrysse, Z.~Yu, and S.~Fan, \enquote{Planar metallic
		nanoscale slit lenses for angle compensation,} Appl. Phys. Lett. \textbf{95},
	071112 (2009).
	
	\bibitem{Ishii11}
	S.~Ishii, A.~V. Kildishev, V.~M. Shalaev, K.-P. Chen, and V.~P. Drachev,
	\enquote{Metal nanoslit lenses with polarization-selective design,} Opt.
	Lett. \textbf{36}, 451--453 (2011).
	
	\bibitem{Arbabi15}
	A.~Arbabi, Y.~Horie, A.~J. Ball, M.~Bagheri, and A.~Faraon,
	\enquote{Subwavelength-thick lenses with high numerical apertures and large
		efficiency based on high-contrast transmitarrays,} Nat. Commun. \textbf{6},
	7069 (2015).
	
	\bibitem{Byrnes16}
	S.~J. Byrnes, A.~Lenef, F.~Aieta, and F.~Capasso, \enquote{Designing large,
		high-efficiency, high-numerical aperture, transmissive meta-lenses for
		visible light,} Opt. Express \textbf{24}, 5110--5124 (2016).
	
	\bibitem{Aieta13}
	F.~Aieta, P.~Genevet, M.~Kats, and F.~Capasso, \enquote{Aberrations of flat
		lenses and aplanatic metasurfaces,} Opt. Express \textbf{21}, 31530--31539
	(2013).
	
	\bibitem{Rho10}
	J.~Rho, Z.~Ye, Y.~Xiong, X.~Yin, Z.~Liu, H.~Choi, G.~Bartal, and X.~Zhang,
	\enquote{Spherical hyperlens for two-dimensional sub-diffractional imaging at
		visible frequencies,} Nat. Commun. \textbf{1}, 143 (2010).
	
	\bibitem{Barnakov11}
	Y.~A. Barnakov, N.~Kiriy, P.~Black, H.~Li, A.~V. Yakim, L.~Gu, M.~Mayy, E.~E.
	Narimanov, and M.~A. Noginov, \enquote{Toward curvilinear metamaterials based
		on silver-filled alumina templates,} Opt. Mater. Express \textbf{1},
	1061--1064 (2011).
	
	\bibitem{Lu12}
	D.~Lu and Z.~Liu, \enquote{Hyperlenses and metalenses for far-field
		super-resolution imaging,} Nat. Commun. \textbf{3}, 1205 (2012).
	
	\bibitem{Naserpour15}
	M.~Naserpour, C.~J. Zapata-Rodr\'{\i}guez, A.~Zakery, and J.~J. Miret,
	\enquote{Highly localized accelerating beams using nano-scale metallic
		gratings,} Opt. Commun. \textbf{334}, 79--84 (2015).
	
	\bibitem{Naserpour15b}
	M.~Naserpour, C.~J. Zapata-Rodr\'{\i}guez, A.~Zakery,
	C.~D\'{\i}az-Avi{\~n}\'{o}, and J.~J. Miret, \enquote{Accelerating wide-angle
		converging waves in the near field,} J. Opt. \textbf{17}, 015602 (2015).
	
	\bibitem{Naserpour15c}
	M.~Naserpour, C.~J. Zapata-Rodr\'{\i}guez, A.~Zakery, and J.~J. Miret,
	\enquote{Light capsules shaped by curvilinear meta-surfaces,} Appl. Phys. B
	pp. 1--6 (2015).
	
	\bibitem{Huang15}
	C.~ping Huang, \enquote{Efficient and broadband polarization conversion with
		the coupled metasurfaces,} Opt. Express \textbf{23}, 32015--32024 (2015).
	
	\bibitem{Yi14}
	X.~Yi, X.~Ling, Z.~Zhang, Y.~Li, X.~Zhou, Y.~Liu, S.~Chen, H.~Luo, and S.~Wen,
	\enquote{Generation of cylindrical vector vortex beams by two cascaded
		metasurfaces,} Opt. Express \textbf{22}, 17207--17215 (2014).
	
	\bibitem{Pfeiffer13}
	C.~Pfeiffer and A.~Grbic, \enquote{Cascaded metasurfaces for complete phase and
		polarization control,} Appl. Phys. Lett. \textbf{102}, 231116 (2013).
	
	\bibitem{Torres15}
	V.~Torres, B.~Orazbayev, V.~Pacheco-Pe{\~n}a, J.~Teniente, M.~Beruete,
	M.~Navarro-C{\'\i}a, M.~Sorolla~Ayza, and N.~Engheta, \enquote{Experimental
		demonstration of a millimeter-wave metallic {ENZ} lens based on the energy
		squeezing principle,} IEEE Trans. Antennas Propag. \textbf{63}, 231--239
	(2015).
	
	\bibitem{Naserpour15d}
	M.~Naserpour, C.~J. Zapata-Rodr\'{\i}guez, C.~D\'{\i}az-Avi{\~n}\'{o},
	M.~Hashemi, and J.~J. Miret, \enquote{Ultrathin high-index metasurfaces for
		shaping focused beams,} Appl. Opt. pp. 7586--7591 (2015).
	
	\bibitem{Lin14}
	D.~Lin, P.~Fan, E.~Hasman, and M.~L. Brongersma, \enquote{Dielectric gradient
		metasurface optical elements,} Science \textbf{345}, 298--302 (2014).
	
	\bibitem{Qin16}
	F.~Qin, L.~Ding, L.~Zhang, F.~Monticone, C.~C. Chum, J.~Deng, S.~Mei, Y.~Li,
	J.~Teng, M.~Hong \emph{et~al.}, \enquote{Hybrid bilayer plasmonic metasurface
		efficiently manipulates visible light,} Sci. Adv. \textbf{2}, e1501168
	(2016).
	
	\bibitem{Rytov56}
	S.~M. Rytov, \enquote{Electromagnetic properties of a finely stratified
		medium,} Sov. Phys. JETP \textbf{2}, 466--475 (1956).
	
	\bibitem{Popov07}
	E.~Popov and S.~Enoch, \enquote{Mystery of the double limit in homogenization
		of finitely or perfectly conducting periodic structures,} Opt. Lett.
	\textbf{32}, 3441--3443 (2007).
	
	\bibitem{Yeh88}
	P.~Yeh, \emph{Optical waves in layered media} (Wiley, New York, 1988).
	
	\bibitem{Elser07}
	J.~Elser, V.~A. Podolskiy, I.~Salakhutdinov, and I.~Avrutsky, \enquote{Nonlocal
		effects in effective-medium response of nanolayered metamaterials,} Appl.
	Phys. Lett. \textbf{90}, 191109 (2007).
	
	\bibitem{Martin01}
	L.~Mart\'{\i}n-Moreno, F.~J. Garc\'{\i}a-Vidal, H.~J. Lezec, K.~M. Pellerin,
	T.~Thio, J.~B. Pendry, and T.~W. Ebbesen, \enquote{Theory of extraordinary
		optical transmission through subwavelength hole arrays,} Phys. Rev. Lett.
	\textbf{86}, 1114--1117 (2001).
	
	\bibitem{Chen10}
	W.~Chen, M.~D. Thoreson, S.~Ishii, A.~V. Kildishev, and V.~M. Shalaev,
	\enquote{Ultra-thin ultra-smooth and low-loss silver films on a germanium
		wetting layer,} Opt. Express \textbf{18}, 5124--5134 (2010).
	
	\bibitem{Diaz16}
	C.~D\'{\i}az-Avi{\~n}\'o, D.~Pastor, C.~J. Zapata-Rodr\'{\i}guez, M.~Naserpour,
	R.~Koty\'nski, and J.~J. Miret, \enquote{Some considerations on the
		transmissivity of trirefringent metamaterials,} J. Opt. Soc. Am. B
	\textbf{33}, 116--125 (2016).
	
	\bibitem{Shin09}
	J.~Shin, J.-T. Shen, and S.~Fan, \enquote{Three-dimensional metamaterials with
		an ultrahigh effective refractive index over a broad bandwidth,} Appl. Phys.
	Lett. \textbf{102}, 093903 (2009).
	
	\bibitem{Choi11}
	M.~Choi, S.~H. Lee, Y.~Kim, S.~B. Kang, J.~Shin, M.~H. Kwak, K.-Y. Kang, Y.-H.
	Lee, N.~Park, and B.~Min, \enquote{A terahertz metamaterial with unnaturally
		high refractive index,} Nature \textbf{470}, 369--373 (2011).
	
	\bibitem{Lee15}
	S.~Lee, \enquote{Colloidal superlattices for unnaturally high-index
		metamaterials at broadband optical frequencies,} Opt. Express \textbf{23},
	28170--28181 (2015).
	
	\bibitem{Huang16}
	Y.~Huang, Q.~Zhao, S.~K. Kalyoncu, R.~Torun, and O.~Boyraz,
	\enquote{Silicon-on-sapphire {mid-IR} wavefront engineering by using
		subwavelength grating metasurfaces,} J. Opt. Soc. Am. B \textbf{33}, 189--194
	(2016).
	
	\bibitem{Mudry10}
	E.~Mudry, E.~L. Moal, P.~Ferrand, P.~C. Chaumet, and A.~Sentenac,
	\enquote{Isotropic diffraction-limited focusing using a single objective
		lens,} Phys. Rev. Lett. \textbf{105}, 203903 (2010).
	
	\bibitem{Hashemi16}
	M.~Hashemi, A.~Moazami, M.~Naserpour, and C.~J. Zapata-Rodr\'{\i}guez,
	\enquote{A broadband multifocal metalens in the terahertz frequency range,}
	Opt. Commun. \textbf{370}, 306--310 (2016).
	
\end{thebibliography}

\end{document}